# High Transmission Bit Rate of A thermal Arrayed Waveguide Grating (AWG) Module in Passive Optical Networks

Abd El–Naser A. Mohammed[1], Ahmed Nabih Zaki Rashed[2*],
Gaber E. S. M. El-Abyad[3] and Abd El–Fattah A. Saad[4]

[1,2,3,4]Electronics and Electrical Communication Engineering Department
Faculty of Electronic Engineering, Menouf 32951, Menoufia University, EGYPT
[1]E-mail: abd_elnaser6@yahoo.com, [2*]E-mail: ahmed_733@yahoo.com

Tel.: +2 048-3660-617, Fax: +2 048-3660-617

*Abstract*—In the present paper, high transmission bit rate of a thermal arrayed waveguide grating (AWG) which is composed of lithium niobate ($LiNbO_3$)/polymethyl metha acrylate (PMMA) hybrid materials on a silicon substrate in Passive Optical Networks (PONs) has parametrically analyzed and investigated over wide range of the affecting parameters. We have theoretically investigated the temperature dependent wavelength shift of the arrayed waveguide grating (AWG) depends on the refractive-indices of the materials and the size of the waveguide. A thermalization of the AWG can be realized by selecting proper values of the material and structural parameters of the device. Moreover, we have analyzed the data transmission bit rate of a thermal AWG in passsive optical networks (PONs) based on Maximum Time Division Multiplexing (MTDM) technique.

*Keywords*−*PONs; Arrayed waveguide gratings (AWGs); integrated optics; optical planar waveguide; optical fiber communications; MTDM technique.*

## I. INTRODUCTION

With the explosive growth of end user demand for higher bandwidth, various types of passive optical networks (PONs) have been proposed. PON can be roughly divided into two categories such as time-division-multiplexing (TDM) and ultra wide wavelength-division-multiplexing (UW-WDM) methods [1]. Compared with TDM-PONs, WDM-PON systems allocate a separate wavelength to each subscriber, enabling the delivery of dedicated bandwidth per optical network unit (ONU). Moreover, this virtual point-to-point connection enables a large guaranteed bandwidth, protocol transparency, high quality of service, excellent security, bit-rate independence, and easy upgradeability. Especially, recent good progress on athermal arrayed waveguide grating (AWG) and cost-effective colorless ONUs [2] has empowered WDM-PON as an optimum solution for the access network. However, fiber link failure from the optical line terminal (OLT) to the ONU leads to the enormous loss of data. Thus, fault monitoring and network protection are crucial issues in network operators for reliable network. To date [3], many methods have been proposed for network protection. In the ITU-T recommendation on PONs (G.983.1) duplicated network resources such as fiber links or ONUs are required. The periodic and cyclic properties of AWGs are used to interconnect two adjacent ONUs by a piece of fiber. In the recent years, arrayed waveguide gratings (AWGs) have appeared to be one of attractive candidates for high channel count Mux/DeMux devices to process optical signals in a parallel manner. Its low chromatic dispersion [4], typically ±5 ps/nm — ±10 ps/nm, makes it possibly be used for 40 Gbit/s systems. However, it is well known that manufacturing AWGs involves a series of complex production processes and requires bulky facilities [5]. Their cost remains a big issue. Further, the technical complexity leads to low yield and poor performance. The former, no doubt, further increases the production cost while the latter degrades the signal quality and system's performance [6], exhibiting high insertion loss, high channel crosstalk, low channel uniformity, and high polarization dependent loss. More vitally, AWGs require active temperature control in order to stabilize the thermal wavelength drift and temperature-dependent loss variations [7]. Due to its capability to increase the aggregate transmission capacity of a single-strand optical fiber, the arrayed waveguide grating (AWG) multiplexer is considered a key component in the construction of a dense wavelength-division-multiplexing system [8]. However, an AWG made of silica is so sensitive to the ambient temperature that the output wavelength changes by as much as 0.66 nm/ºC [9]. In the present study, a hybrid material waveguide with lithium niobate ($LiNbO_3$) core material and PMMA cladding material is considered as the most attractive a thermal structure because of its resistance to the thermo-optic sensitivity of the materials. First, the principle of the a thermal AWG with $LiNbO_3$/PMMA hybrid materials is described, and the relative formulas are derived for analyzing the temperature dependence of the AWG. Second, the theoretical analysis of the data transmission bit rate of a thermal AWG in PONs based on MTDM technique . Finally, a conclusion is reached based on the analysis and general discussion.

13

## II. A THERMAL AWG MODULE IN PONS ARCHITECTURE MODEL

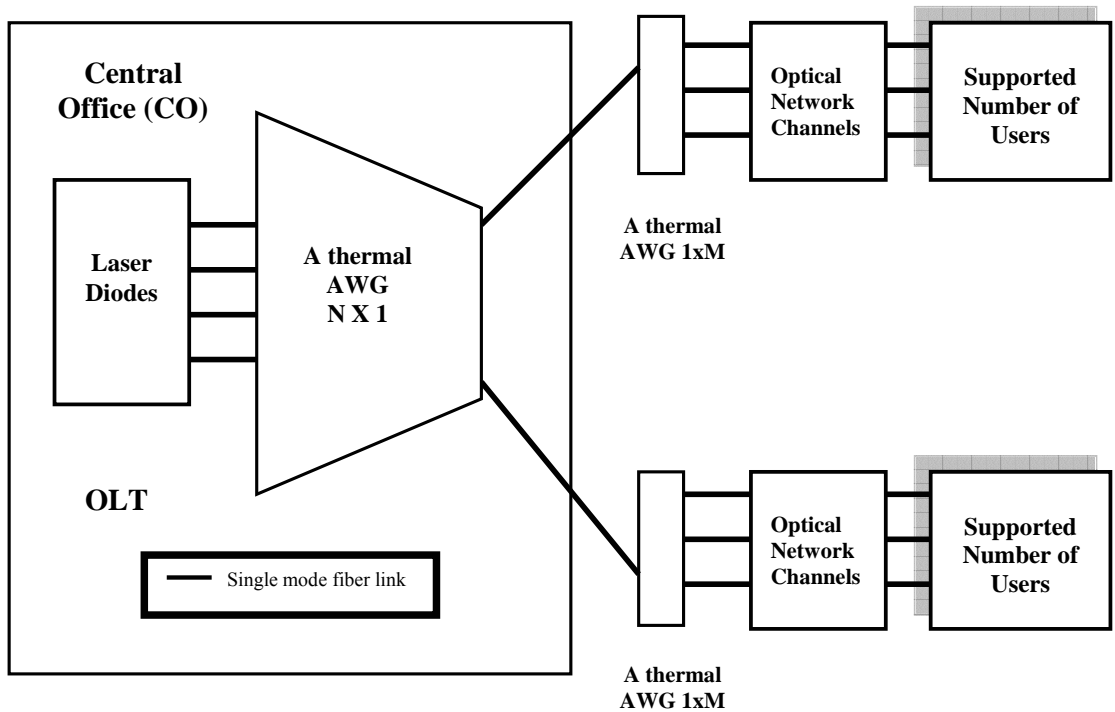

Fig. 1. Simplified Passive Optical network architecture model.

The network architecture is shown in Fig. 1. It is based on two cascaded Athermal arrayed waveguide gratings (AWGs). The first stage is an Nx1a thermal AWG located at the optical line terminal (OLT) or central office (CO). The functionality of this a thermal AWG is to route optical signals generated by the OLT laser diodes stack to each of the network branches to which the OLT will serve [4]. The second stage is a 1× M a thermal AWG located at the remote node. Its task is to demultiplex the M incoming wavelengths to each of the output ports, which connect to optical network channels and then to the number of supported users. The entire network routing intelligence is located at the CO in order to provide easy upgradeability and easy integration with the backbone. The two cascade a thermal AWGs are connected to each other by the single mode optical fiber cables [5].

## III. Features Of A Thermal Arrayed Waveguide Grating Module

Arrayed waveguide grating (AWG) which handles the function of wavelength multiplexer/demultiplexer is extensively used in configuring optical communication networks that are becoming more diversified. Since the transmission wavelength of an AWG is temperature dependent, it was a common practice to control its temperature using heaters or Peltier elements. The AWG consists of input waveguides, arrayed waveguide, slab waveguide and output waveguides, constituting a diffraction grating that takes advantage of the optical path difference in the arrayed waveguide. Figure 2 shows a schematic view of AWG circuitry. As ambient temperature changes, the phase front at each wavelength generated by the arrayed waveguide will tilt due to the change in the refractive index of the optical waveguide and the linear expansion of the silicon substrate, causing a shift of the focusing point on the output waveguide within the slab waveguide [10].

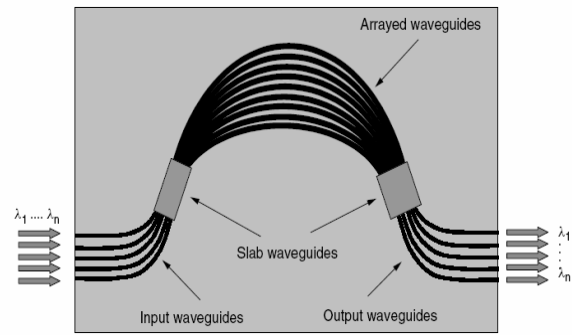

Fig. 2. Sechamtic view of a thermal AWG.

As shown in Fig. 3, the thermal arrayed waveguide grating (AWG) with cross-sections is designed as square shape with the core width a, of lithium niobate ($LiNbO_3$) material and PMMA polymer overcladding and undercladding material on a silicon substrate.

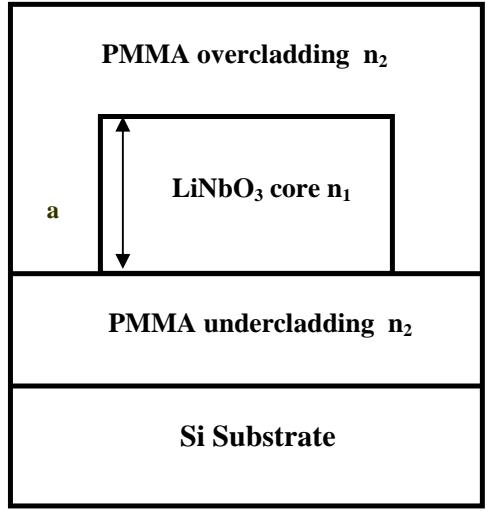

Fig. 3. A structure view of cross-section and refractive- index profile of hybrid materials LiNbO$_3$/PMMA.

## IV. REFRACTIVE-INDEX OF HYBRID MATERIALS

### IV. 1. Lithium Niobate (LiNbO$_3$) Core Material

The investigation of both the thermal and spectral variations of the waveguide refractive index (n) require Sellmeier equation. The set of parameters required to completely characterize the temperature dependence of the refractive-index ($n_1$) is given below, Sellmeier equation is under the form [11]:

$$n_1^2 = A_1 + A_2 H + \frac{A_3 + A_4 H}{\lambda^2 - (A_5 + A_6 H)^2} + \frac{A_7 + A_8 H}{\lambda^2 - A_9^2} - A_{10}\lambda^2 \quad (1)$$

where $\lambda$ is the optical wavelength in μm and $H = T^2 - T_0^2$. T is the temperature of the material, C, and $T_0$ is the reference temperature and is considered as 27 C. The set of parameters of Sellmeier equation coefficients, LiNbO$_3$, are recast and dimensionally adjusted as below [11]: $A_1$=5.35583, $A_2$=4.629 x 10$^{-7}$, $A_3$=0.100473, $A_4$=3.862 x 10$^{-8}$, $A_5$=0.20692, $A_6$= -0.89 x 10$^{-8}$, $A_7$=100, $A_8$=2.657 x 10$^{-5}$, $A_9$=11.34927, and $A_{10}$=0.01533.

Equation (1) can be simplified as the following expression:

$$n_1^2 = A_{12} + \frac{A_{34}}{\lambda^2 - A_{56}^2} + \frac{A_{78}}{\lambda^2 - A_9^2} - A_{10}\lambda^2 \quad (2)$$

where: $A_{12}=A_1+A_2 H$, $A_{34}=A_3+A_4 H$, $A_{56}=A_5+A_6 H$, and $A_{78}=A_7+A_8 H$.

Then, the differentiation of Eq. (2) w. r. t $\lambda$ which gives:

$$\frac{dn_1}{d\lambda} = \left(\frac{-\lambda}{n_1}\right)\left[\frac{A_{34}}{(\lambda^2 - A_{56}^2)^2} + \frac{A_{78}}{(\lambda^2 - A_9^2)^2} + A_{10}\right] \quad (3)$$

In the same way, the second differentiation w. r. t $\lambda$ yields:

$$\frac{d^2 n_1}{d\lambda^2} = -\frac{1}{n_1}\left[\frac{A_{34}[(\lambda^2 - A_{56}^2) - 4\lambda^2]}{(\lambda^2 - A_{56}^2)^3} + \frac{A_{78}[(\lambda^2 - A_9^2) - 4\lambda^2]}{(\lambda^2 - A_9^2)^3} + A_{10}\right] \quad (4)$$

Also, the differentiation w. r. t T gives:

$$\frac{dn_1}{dT} = \left(\frac{T}{n_1}\right)\left[A_2 + \frac{(\lambda^2 - A_{56}^2)A_4 + 2 A_6 A_{56} A_{34}}{(\lambda^2 - A_{56}^2)^2} + \frac{A_8}{(\lambda^2 - A_9^2)}\right] \quad (5)$$

### IV. 2. PMMA Polymer Cladding Material

The Sellmeier equation of the refractive-index is expressed as the following [12]:

$$n_2^2 = 1 + \frac{C_1 \lambda^2}{\lambda^2 - C_2^2} + \frac{C_3 \lambda^2}{\lambda^2 - C_4^2} + \frac{C_5 \lambda^2}{\lambda^2 - C_6^2} \quad (6)$$

The parameters of Sellmeier equation coefficients, PMMA, as a function of temperature [12]: $C_1$=0.4963, $C_2$=0.07180 (T/T$_0$), $C_3$=0.6965, $C_4$=0.1174 (T/T$_0$), $C_5$=0.3223, and $C_6$=9.237.

Then the differentiation of Eq. (6) w. r. t T yields:

$$\frac{dn_2}{dT} = \frac{\lambda^2 (0.0718)}{n_2 T_0}\left[\frac{C_1 C_2}{(\lambda^2 - C_2^2)^2} + \frac{1.635 C_3 C_4}{(\lambda^2 - C_4^2)^2}\right] \quad (7)$$

## V. THEORETICAL MODEL ANALYSIS

### V. 1. Model of A thermal Arrayed Waveguide Grating (AWG)

We present the a thermal condition and the relative formulas of LiNbO$_3$/PMMA hybrid materials AWG on a silicon substrate. The temperature dependence of AWG center wavelength is expressed as [13, 14].

$$\frac{d\lambda_c}{dT} = \frac{\lambda_c}{n_c}\left(\frac{dn_c}{dT} + n_c \alpha_{sub}\right) \quad (8)$$

where T is the ambient temperature, C, $\lambda_c$ is the center wavelength of the arrayed waveguide grating, μm, $n_c$ is the effective refractive-index of the arrayed waveguide grating, $\alpha_{sub}$ is the coefficient of thermal expansion of the Si substrate, and $\frac{dn_c}{dT}$ is the thermo-optic (TO) coefficient of the waveguide. By integrating Eq. (8), we can obtain the following expression:

$$\lambda_c = C n_c e^{(\alpha_{sub} T)} \quad (9)$$

where C is an integrating constant. Assume that $\lambda_c = \lambda_0$, and $n_c = n_{c0}$ when $T=T_0$ at room temperature, we can determine C as the following:

$$C = \frac{\lambda_0}{n_{c0}} e^{(-\alpha_{sub} T_0)} \quad (10)$$

Substituting from Eq. (10) into Eq. (9), we can obtain:

$$\lambda_c = \frac{\lambda_0 n_c}{n_{c0}} e^{[\alpha_{sub}(T-T_0)]} \qquad (11)$$

From Eq. (11) we obtain the central wavelength shift caused by the temperature variation as:

$$\Delta\lambda = \lambda_c - \lambda_0 = \frac{\lambda_0}{n_{c0}}\left[ n_c e^{(\alpha_{sub}(T-T_0))} - n_{co} \right] \qquad (12)$$

Taking $\Delta\lambda = 0$, from Eq. (12) we can obtain the a thermal condition of the AWG as:

$$\alpha_{sub}(T-T_0) = \ln\left(\frac{n_c}{n_{c0}}\right) \qquad (13)$$

Then by differentiating Eq. (13), the a thermal condition of the AWG can also be expressed in another form as the following [14]:

$$\frac{dn_c}{dT} = -\alpha_{sub} n_c \qquad (14)$$

The effective refractive index of the arrayed waveguide grating (AWG) is given by [15]:

$$n_c = \frac{\beta}{k} = \frac{k\left[(n_1^2 - n_2^2)b + n_2^2\right]}{k} = (n_1^2 - n_2^2)b + n_2^2 \quad , \quad (15)$$

where $\beta$ is the propagation constant of the fundamental mode, k is the wave number, and b is the normalized propagation constant and is given by [15]:

$$b(V) = \left(1.1428 - \frac{0.9660}{V}\right)^2 , \qquad (16)$$

where V is the normalized frequency. For single mode step index optical fiber waveguide, the cut-off normalized is approximately $V = V_c = 2.405$, and by substituting in Eq. (16) we can get the normalized propagation constant b at the cut-off normalized frequency approximately $b \approx 0.5$, and then by substituting in Eq. (15) we can obtain:

$$n_c = \frac{1}{2}\left(n_1^2 + n_2^2\right) , \qquad (17)$$

By taking the square root of Eq. (17) yields:

$$\sqrt{n_c} = 0.7\left(n_1^2 + n_2^2\right)^{1/2} , \qquad (18)$$

The cut-off normalized frequency for single mode step index optical fiber waveguide is given by the following expression [15]:

$$V_c = \frac{2\pi a}{\lambda_{cut-off}}\left(n_1^2 - n_2^2\right)^{1/2} , \qquad (19)$$

Assume that the cut-off wavelength is equal to the central wavelength to transfer the fundamental modes only, that is $\lambda_{cut-off} = \lambda_c$. Eq. (19) can be expressed in another form as follows:

$$\lambda_c = \frac{1.4\pi a \left(n_1^4 - n_2^4\right)^{1/2}}{2.405 \sqrt{n_c}} , \qquad (20)$$

Equation (20) can be simplified as follows:

$$\lambda_c = \frac{1.83 a \left(n_1^4 - n_2^4\right)^{1/2}}{\sqrt{n_c}} , \qquad (21)$$

Equation (21) can be expressed in another form as follows:

$$n_c = \frac{3.35 a^2 \left(n_1^4 - n_2^4\right)}{\lambda_c^2} , \qquad (22)$$

By substituting from Eq. (22) into Eq. (14) yields:

$$\frac{dn_c}{dT} = -\frac{3.35 \alpha_{sub} a^2 \left(n_1^4 - n_2^4\right)}{\lambda_c^2} . \qquad (23)$$

The effective refractive index $n_c$ is dependent on the refractive indices of the materials and on the size and shape of the waveguide, then by selecting proper materials and structural parameters of the waveguide to satisfy Eq. (23), an a thermal arrayed waveguide grating (AWG) can be designed.

### V. 2. Theoretical Model Analysis of High Data Transmission Bit Rate

The total B.W is based on the total chromatic dispersion coefficient $D_t$ [16], where:

$$D_t = D_m + D_w \qquad (24)$$

where $D_m$ is the material dispersion coefficient in sec/m², and $D_w$ is the waveguide dispersion coefficient in sec/m². Both $D_m$, $D_w$ are given by [17] (for the fundamental mode):

$$D_m = -\frac{\lambda}{C}\left(\frac{d^2 n_1}{d\lambda^2}\right), \quad \text{sec/m}^2 \qquad (25)$$

$$D_w = -\left(\frac{n_2}{C n_1}\frac{\Delta n}{\lambda}\right) Y , \quad \text{sec/m}^2 \qquad (26)$$

where C is the velocity of the light, $3 \times 10^8$ m/sec, $n_1$ is the core refractive-index, $n_2$ is the cladding refractive-index, Y is a function of wavelength, $\lambda$ [17].

The relative refractive-index difference $\Delta n$ is defined as [17]:

$$\Delta n = \frac{n_1^2 - n_2^2}{2n_1^2} \qquad (27)$$

The total pulse broadening due to total chromatic dispersion coefficient $D_t$ is given by [17]:

$$\Delta\tau = D_t L \Delta\lambda , \quad \text{nsec} \qquad (28)$$

where $\Delta\lambda$ is the spectral line-width of the optical source, nm, and L is the length of single-mode fiber waveguide, m.

The maximum time division multiplexing (MTDM) transmission bit rate is given by [17]:

$$B_{rm} = \frac{1}{4\Delta\tau} = \frac{0.25}{\Delta\tau} , \quad \text{Gbit/sec} \qquad (29)$$

The optical signal wavelength span 1 μm ≤ $\lambda_{si}$, optical signal wavelength ≤ 1.65 μm is divided into intervals per link as follows:

$$\Delta\lambda_0 = \frac{\lambda_f - \lambda_i}{N_L} = \frac{0.65}{N_L} , \quad \mu m/link \qquad (30)$$

Then the MTDM bit rates per fiber cable link is given by the following expression:

$$B_{rLink} = \frac{0.25 \times N_{ch}}{\Delta\tau} , \quad Gbit/sec/link \qquad (31)$$

where $N_{ch}$ is the number of optical network channels in the fiber cable link, and $N_L$ is the number of links in the fiber cable core and is up to 24 links/core.

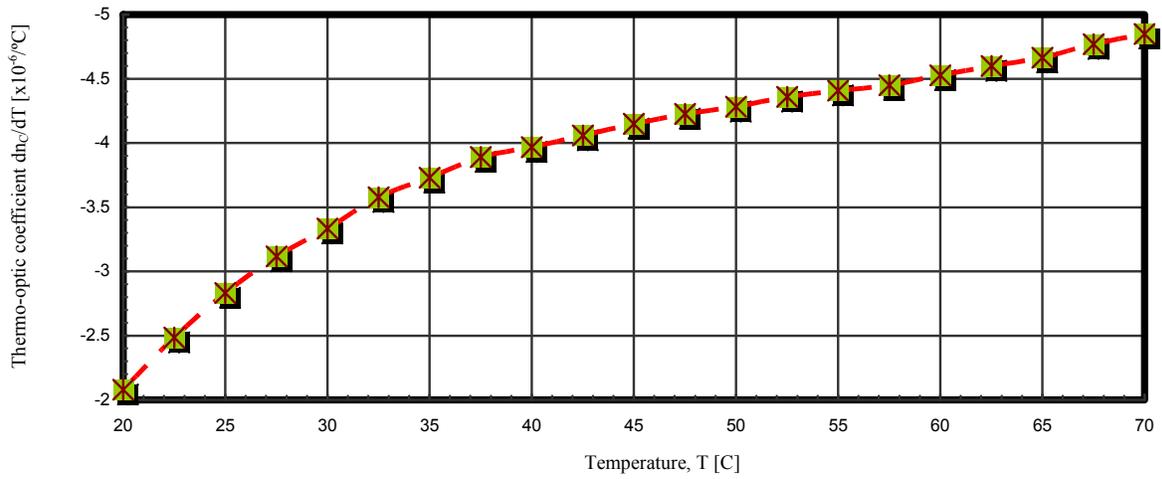

Fig. 4. Variation of thermo-optic (TO) coefficient versus temperature when $n_1$=2.33, $n_2$=1.52, a= 5 μm.

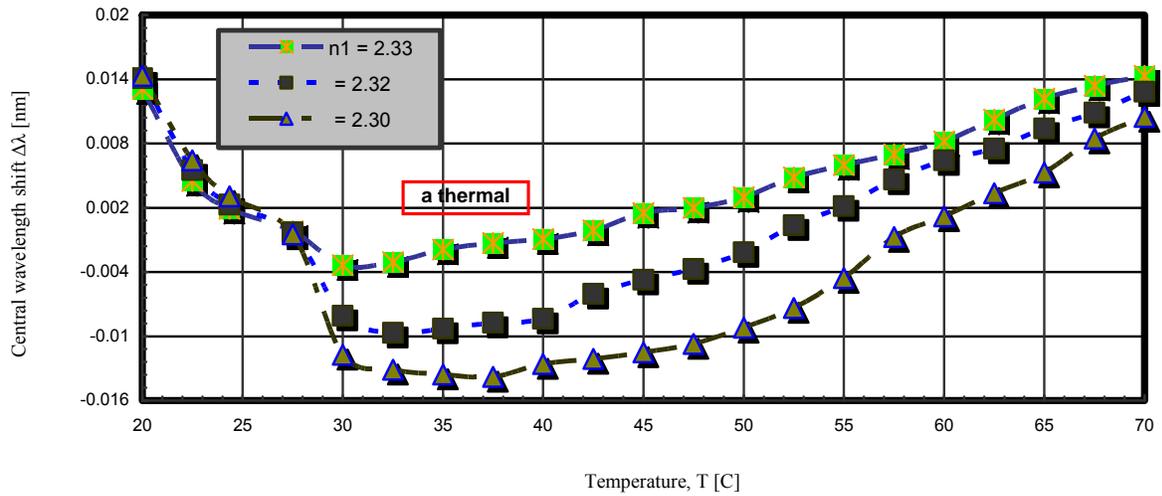

Fig. 5. Variation of the central wavelength shift versus temperature for different core refractive-indices.

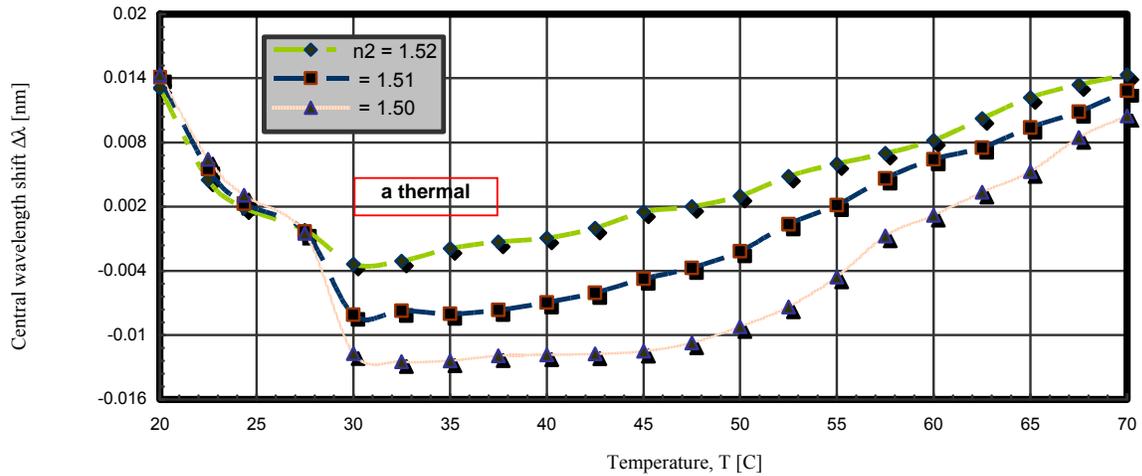

Fig. 6. Variation of the central wavelength shift versus temperature for different cladding refractive-indices.

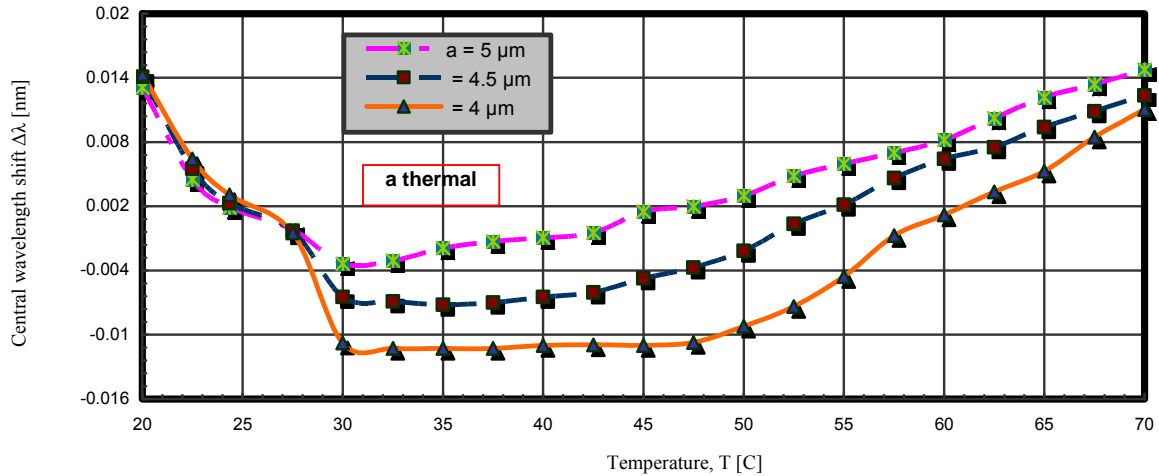

Fig. 7. Variation of the central wavelength shift versus temperature for different core width of a thermal AWG.

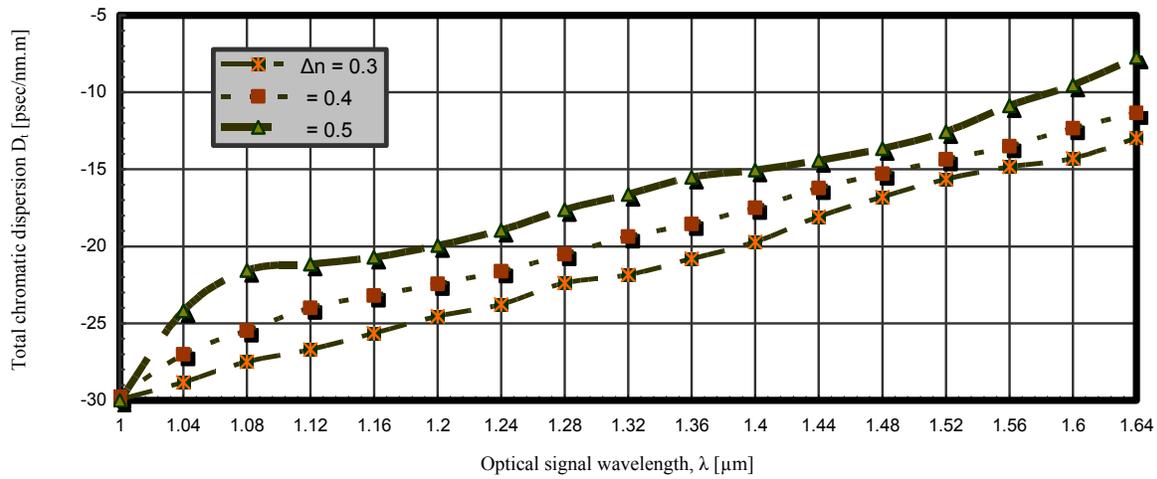

Fig. 8. Total chromatic dispersion coefficient versus optical signal wavelength at the assumed set of parameters.

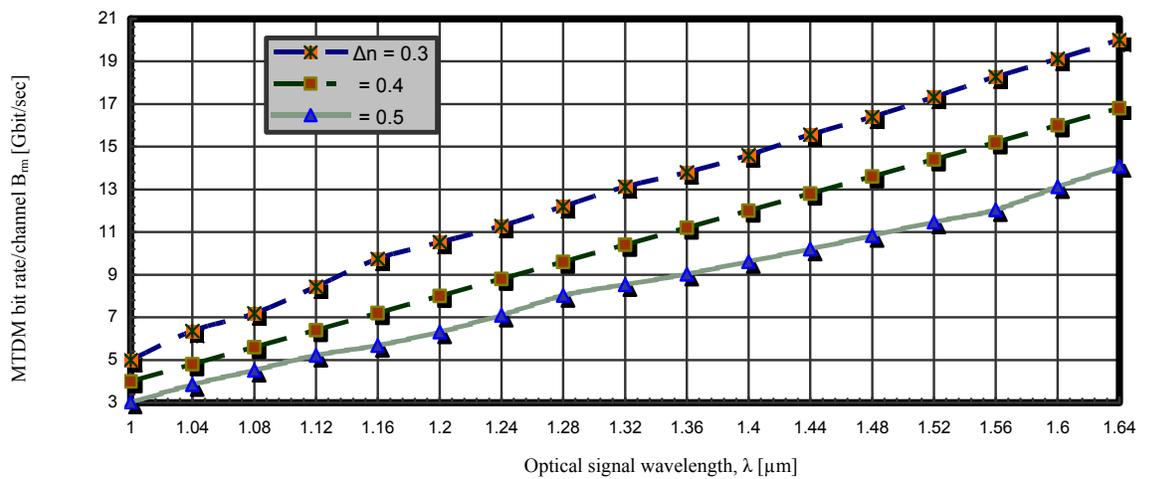

Fig. 9. MTDM transmission bit rate/channel versus optical signal wavelength a the assumed set of parameters.

**18**

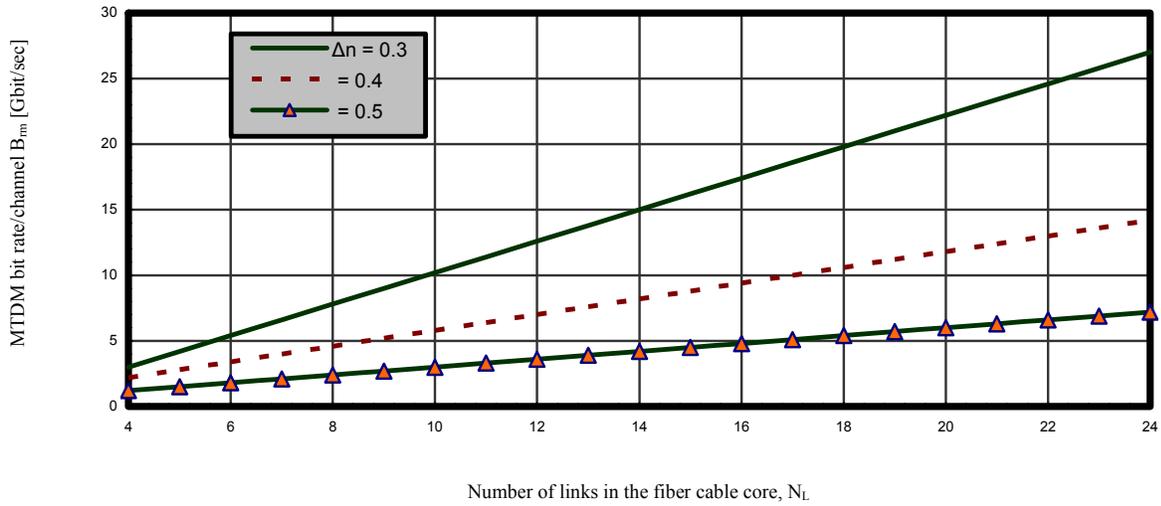

Fig. 10. MTDM transmission bit rate/channel versus number of links in the fiber cable core at the assumed set of parameters.

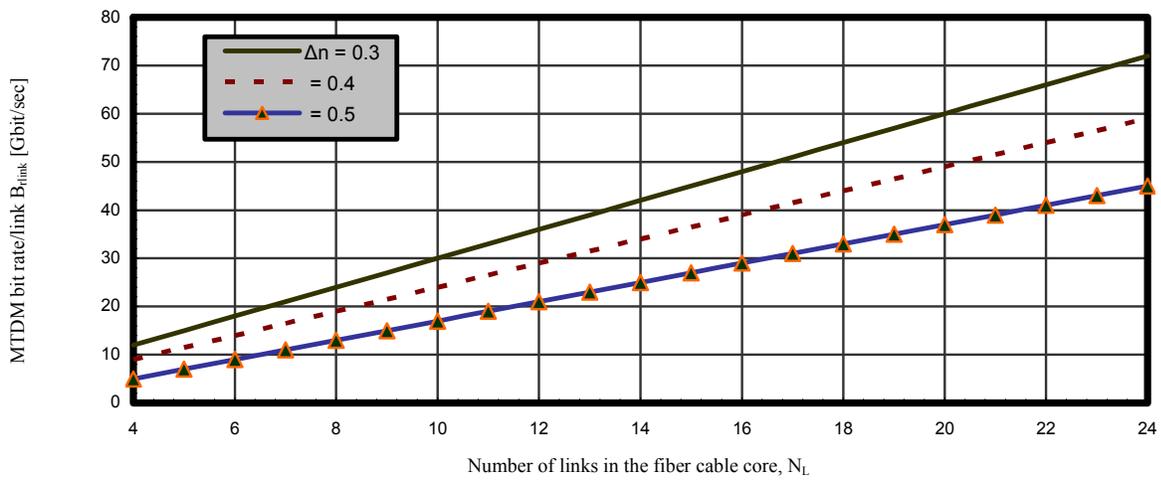

Fig. 11. MTDM transmission bit rate/link versus number of links in the fiber cable core at the assumed set of parameters.

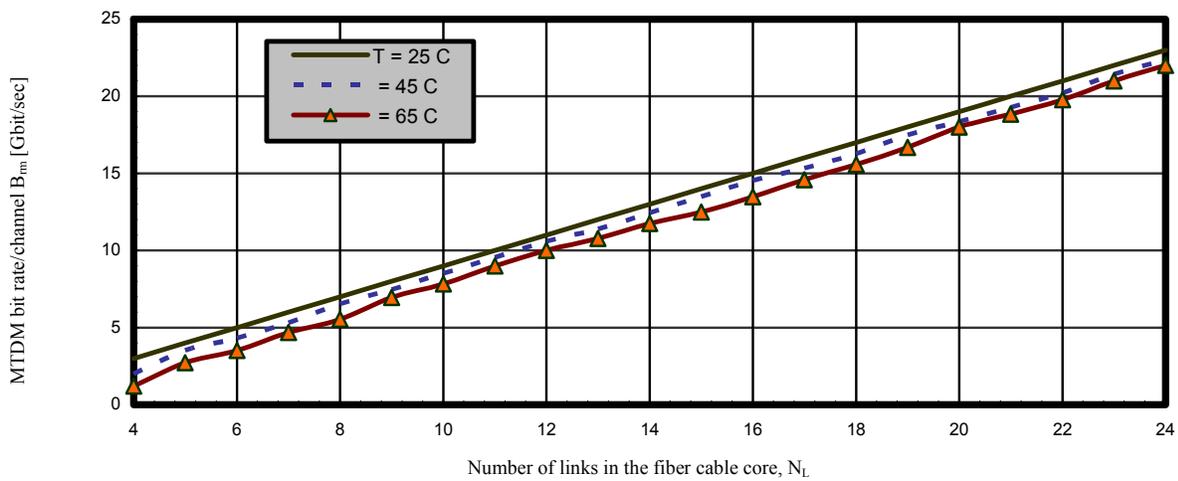

Fig. 12. MTDM transmission bit rate/channel versus number of links in the fiber cable core at the assumed set of parameters.

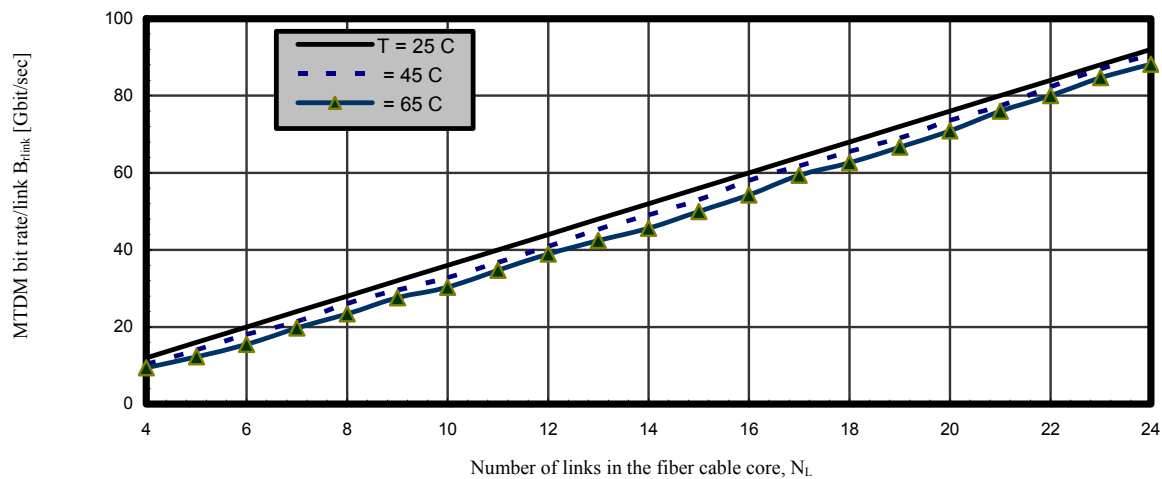

Fig. 13. MTDM transmission bit rate/link versus number of links in the fiber cable core at the assumed set of parameters.

## VI. RESULTS AND DISCUSSIONS

The center wavelength at room temperature $T_0$=27 C is selected to be $\lambda_0$= 1.550918 μm, which is one of the standard wavelengths recommended by the International Telecommunication Union (ITU) [16]. This AWG device is made on the silicon substrate have a coefficient of thermal expansion of $\alpha_{sub}$= 2.63x10$^{-6}$/C [16]. Because the environmental temperature of an AWG is usually changed from 20 C t0 70 C. We discuss the central wavelength shift Δλ in this range of temperature variation. The subsequent relations between wavelength shift, and refractive-indices of core, and cladding $n_1$, $n_2$ as well as the core width **a** are discussed as follows. Also, we discuss the maximum transmission bit rate of AWG device model in the operating wavelength range from 1 μm to 1.64 μm as follows.

1) Figure 4 has indicated the dependence of the thermo-optic (TO) coefficient $dn/dT$ on the temperature T. We can find that $dn/dT$ is not constant with the variation of temperature which nonlinearly increases as temperature increases. Therefore, this behavior of $dn/dT$ will obviously affect the shifts of the central wavelength caused by the variation of temperature.
2) As shown in Figs. (5-7) have demonstrated the dependence of the central wavelength shift Δλ on the refractive-indices of the core, and cladding $n_1$, $n_2$ as well as the core width a for the designed a thermal hybrid material AWG, which are calculated from Eq. (12). We can find that there exists an optimal operation condition of the AWG, which should guarantee the central wavelength shift Δλ to be small enough in a sufficiently large range of the temperature variation. To be precise, when we select $n_1$= 2.33, $n_2$=1.52, and a= 5 μm, the central wavelength shift is within the range of 0.012 ~ 0.015 nm as the temperature increases from 20 C to 70 C. In this case we can presume that the a thermalization is realized in the designed AWG.
3) Figure 8 has demonstrated the variation of the total chromatic dispersion $D_t$ against the variation of optical signal wavelength within the range from 1 μm to 1.64 μm for different relative refractive-index difference Δn. We can find that the smaller Δn, the smaller $|D_t|$ within the same variation of the optical signal wavelength.
4) As shown in Fig. (9), the variation of the MTDM transmission bit rate, against the variation of optical signal wavelength within the range from 1 μm to 1.64 μm for different relative refractive-index difference Δn. We can find that the smaller Δn, the larger the bit rate within the same variation of the optical signal wavelength.
5) Figures (10, 11) have indicated that as the number of links in the fiber cable core increse, MTDM bit rate either Per link or Per channel increases at the same relative refractive index difference Δn. While the smaller of Δn, the higher of bit rates either per link or per channel at the same number of links in the fiber cable core.
6) Figures (12, 13) have indicated that as the number of links in the fiber cable core increse, MTDM bit rate either Per link or Per channel increases at the same ambient temperature. While the smaller of T, the slightly higher of bit rates either per link or per channel at the same number of links in the fiber cable core.

## VII. CONCLUSIONS

In a summary, we have presented a novel technique for theoretical simulation and optimum design of the a thermal AWG with LiNbO$_3$/PMMA hybrid materials. By selecting the proper values of the refractive-indices of the materials and the core size of the waveguide, the a thermalization can be realized. To be precise, the central wavelength shifts of the designed a thermal hybrid material AWG only increases to 0.027 nm/C, while that of the conventional silica-based AWG increases to 0.66 nm/C [9], our designed a thermal hybrid material AWG showed a good performance with Δλ = ~ 0.027 nm/C over the temperature range of 20 C to 70 C. Finally, we can conclude that the smaller Δn, and the lower ambient temperature T, The higher transmission bit rate of our a thermal hybrid material AWG model either per link or per channel which is appropriate for passive optical networks standards.

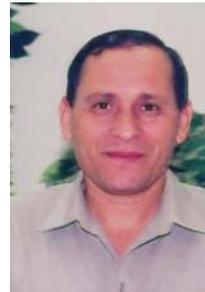

**Abd-Elnaser A. Mohammed**

Received Ph.D degree from the faculty of Electronic Engineering, Menoufia University in 1994. Now, his job career is Assoc. Prof. Dr. in Electronics and Electrical Communication Engineering department. Currently, his field and research interest in the passive optical communication Networks, digital communication systems, and advanced optical communication networks.

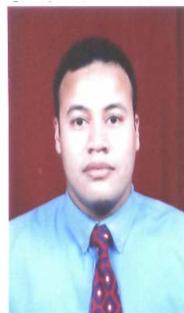

**Ahmed Nabih Zaki Rashed**

was born in Menouf, Menoufia State, Egypt, in 1976. Received the B.Sc. and M.Sc. scientific degrees in the Electronics and Electrical Communication Engineering Department from Faculty of Electronic Engineering, Menoufia University in 1999 and 2005, respectively. Currently, his field interest and working toward the Ph.D degree in Active and Passive Optical Networks (PONs). His research mainly focuses on the transmission data rate of optical networks.